\documentclass[11pt]{article}

\usepackage[left=1.25in,top=1.0in,right=1.25in,bottom=1.0in]{geometry}
\usepackage{amsfonts}
\usepackage{amssymb}
\usepackage{graphicx}
\usepackage{setspace}
\usepackage[round]{natbib}
\usepackage{amsmath}
\usepackage{amssymb}
\usepackage{amsbsy}
\usepackage{amsthm}
\usepackage{paralist}

\usepackage{mathtools}
\usepackage{cases}
\usepackage{algorithm}
\usepackage{algpseudocode}

\newcommand{\nc}{\newcommand}

\newcommand{\N}{\mbox{N}}

\newcommand{\Ga}{\mbox{Ga}}

\newcommand{\bbeta}{\beta}

\newcommand{\by}{y}

\nc{\bone}{{\bf 1}}

\nc{\bX}{{X}}
\nc{\Li}{\hat\Lambda^{-1(g)}}
\nc{\Oi}{\hat\Omega^{-1(g)}}
\nc{\diag}[1]{\text{diag}\left(#1\right)}
\nc{\Siginv}{\Sigma^{-1}}
\nc{\Ominv}{\Omega^{-1}}

\newtheorem{theorem}{Theorem}[section]

\newtheorem{corollary}[theorem]{Corollary}

\title{The Bayesian Bridge}
\author{Nicholas G. Polson\\
\textit{University of Chicago}\footnote{Polson is Professor of Econometrics and Statistics
at the Chicago Booth School of Business. email: ngp@chicagobooth.edu. Scott
is Assistant Professor of Statistics at the University of Texas at Austin.
email: James.Scott@mccombs.utexas.edu.  Windle is a Ph.D student at the University of Texas at Austin.  email: jwindle@ices.utexas.edu.}\\ \\
James G. Scott\\
Jesse Windle \\
\textit{University of Texas at Austin}}
\date{First Version: July 2011\\
This Version: October 2012}

\begin{document}

\begin{spacing}{1.1}

\maketitle
\begin{abstract}
\noindent We propose the Bayesian bridge estimator for regularized regression and classification. Two key mixture representations for the Bayesian bridge model are developed: (1) a scale mixture of normals with respect to an alpha-stable random variable; and (2) a mixture of Bartlett--Fejer kernels (or triangle densities) with respect to a two-component mixture of gamma random variables.  Both lead to MCMC methods for posterior simulation, and these methods turn out to have complementary domains of maximum efficiency.  The first representation is a well known result due to \citet{west:1987}, and is the better choice for collinear design matrices.  The second representation is new, and is more efficient for orthogonal problems, largely because it avoids the need to deal with exponentially tilted stable random variables.  It also provides insight into the multimodality of the joint posterior distribution, a feature of the bridge model that is notably absent under ridge or lasso-type priors \citep{park:casella:2008,hans:2008}.   We prove a theorem that extends this representation to a wider class of densities representable as scale mixtures of betas, and provide an explicit inversion formula for the mixing distribution.  The connections with slice sampling and scale mixtures of normals are explored.  On the practical side, we find that the Bayesian bridge model outperforms its classical cousin in estimation and prediction across a variety of data sets, both simulated and real.  We also show that the MCMC for fitting the bridge model exhibits excellent mixing properties, particularly for the global scale parameter.  This makes for a favorable contrast with analogous MCMC algorithms for other sparse Bayesian models.  All methods described in this paper are implemented in the R package \verb|BayesBridge|.  An extensive set of simulation results are provided in two supplemental files.
\end{abstract}

\newpage

\section{Introduction}

\subsection{Penalized likelihood and the Bayesian bridge}

This paper develops the Bayesian analogue of the bridge estimator in regression, where $\by = X \bbeta + \epsilon$ for some unknown vector $\bbeta = (\beta_1, \ldots, \beta_p)'$.  Given choices of $\alpha \in (0,1]$ and $\nu \in \mathbb{R}^+$, the bridge estimator $\hat{\bbeta}$ is the minimizer of
\begin{equation}
\label{eqn:bridgeobjective}
Q_{\by}(\bbeta) =  \frac{1}{2} || \by - X \bbeta ||^2 + \nu \sum_{j=1}^p | \beta_j |^\alpha \, .
\end{equation}
This bridges a class of shrinkage and selection operators, with the best-subset-selection penalty at one end, and the $\ell^1$ (or lasso) penalty at the other.  An early reference to this class of models can be found in \citet{frank:friedman:1993}, with recent papers focusing on model-selection asymptotics, along with strategies for actually computing the estimator \citep{huang:horowitz:ma:2008,zou:li:2008,mazumder:friedman:hastie:2009}.

Our approach differs from this line of work in adopting a Bayesian perspective on bridge estimation.  Specifically, we treat $p(\bbeta \mid \by) \propto \exp\{- Q_{\by}(\beta)\}$ as a posterior distribution having the minimizer of (\ref{eqn:bridgeobjective}) as its global mode.  This posterior arises in assuming a Gaussian likelihood for $\by$, along with a prior for $\bbeta$ that decomposes as a product of independent exponential-power priors \citep{box:tiao:1973}:
\begin{equation}
\label{eqn:bridgeprior}
p(\bbeta \mid \alpha, \nu) \propto \prod_{j=1}^p \exp( -|\beta_j/\tau|^\alpha ) \; , \; \tau = \nu^{-1/\alpha} \, .
\end{equation}
Rather than minimizing (\ref{eqn:bridgeobjective}), we proceed by constructing a Markov chain having the joint posterior for $\bbeta$ as its stationary distribution. 

\subsection{Relationship with previous work}

Our paper emphasizes several interesting features of the Bayesian approach to bridge estimation.  These features can be grouped into three main categories.

\paragraph{Versus the Bayesian ridge and lasso priors.}

There is a large literature on Bayesian versions of classical estimators related to the exponential-power family, including the ridge \citep{lindley:smith:1972}, lasso \citep{park:casella:2008,hans:2008,hans:2010}, and elastic net \citep{li:lin:2010,hans:2011}.  Yet the bridge penalty has a crucial feature not shared by these other approaches: it is concave over $(0, \infty)$.  From a Bayesian perspective, this implies that the prior for $\bbeta$ has heavier-than-exponential tails.  As a result, when the underlying signal is sparse, and when further regularity conditions are met, the bridge penalty dominates the lasso and ridge according to a classical criterion known as the oracle property \citep{fan:li:2001,huang:horowitz:ma:2008}.  Although the oracle property \textit{per se} is of no particular relevance to a Bayesian treatment of the problem, it does correspond to a feature of certain prior distributions that Bayesians have long found important: the property of yielding a redescending score function for the marginal distribution of $\by$ \citep[e.g.][]{pericchi:smith:1992}.  This property is highly desirable in sparse situations, as it avoids the overshrinkage of large regression coefficients even in the presence of many zeros \citep{Polson:Scott:2010a}.  See \citet{griffin:brown:2010} for a discussion of the Bayesian interpretation of the oracle property.

\paragraph{Versus the classical bridge estimator.}

Both the classical and Bayesian approaches to bridge estimation must confront a significant practical difficulty: exploring and summarizing a multimodal surface in high-dimensional Euclidean space.  In our view, multimodality is one of the strongest arguments for pursuing a full Bayes approach.  For one thing, it is misleading to summarize a multimodal surface in terms of a single point estimate, no matter how appealingly sparse that estimate may be.  Moreover, \citet{mazumder:friedman:hastie:2009} report serious computational difficulties with getting stuck in local modes in attempting to minimize (\ref{eqn:bridgeobjective}).  This is related to the fact that, when $0 < \alpha < 1$, the target function fails to be convex and fast, automatic routines are not available.   Our sampling-based approach, while not immune to the problem of local modes, seems very effective at exploring the whole space.  As Section \ref{sec:DAstrategies} will show, there are very good reasons for expecting this to be the case, based on the structure of the data-augmentation strategy we pursue.  In this respect, MCMC behaves like a simulated annealing algorithm that never cools.

In addition, previous authors have emphasized three other points about penalized-likelihood rules that will echo in the examples we present in Section \ref{sec:examples}.  First, one must choose a penalty parameter $\nu$.  In the classical setting this can be done via cross validation, which usually yields reasonable results.  Yet this ignores uncertainty in the penalty parameter, which may be considerable.  We are able to handle this in a principled way by averaging over uncertainty in the posterior distribution, under some default prior for the global variance component $\tau^2$ \citep[e.g.][]{gelman:2006}.  In the case of the bridge estimator, this logic may also be extended to the concavity parameter $\alpha$, for which even less prior information is typically available.

Second, the minimizer of (\ref{eqn:bridgeobjective}) may produce a sparse estimator, but this estimate is provably suboptimal, in a Bayes-risk sense, with respect to most traditional loss functions.  If, for example, one wishes either to estimate $\bbeta$ or to predict future values of $y$ under squared-error loss, then the optimal solution is the posterior mean, not the mode.  Both \citet{park:casella:2008} and \citet{hans:2008} give realistic examples where the ``Bayesian lasso'' significantly outperforms its classical counterpart, both in prediction and in estimation.  Similar conclusions are reached by \citet{efron:2009} in a parallel context.  Our own examples provide evidence of the practical differences that arise on real data sets---not merely between the mean and the mode, but also between the classical bridge solution and the mode of the joint distribution in the Bayesian model, marginal over over $\tau$ and $\sigma$.  In the cases we study, the Bayesian approach leads to lower risk, often dramatically so.

Third, a fully Bayesian approach can often lead to different substantive conclusions than a traditional penalized-likelihood analysis, particularly regarding which components of $\bbeta$ are important in predicting $\by$.  For example, \citet{hans:2010} produces several examples where the classical lasso estimator aggressively zeroes out components of $\bbeta$ for which, according to a full Bayes analysis, there is quite a large amount of posterior uncertainty regarding their size.  This is echoed in our analysis of the classic diabetes data set \citep[see, e.g.][]{efron:LARS:2004}.  This is not to suggest that one conclusion is right, and the other wrong, in any specific setting---merely that the two conclusions can be quite different, and that practitioners are well served by having both at hand.

\paragraph{Versus other sparsity-inducing priors in Bayesian regression analysis.}

Within the broader class of regularized estimators in high-dimensional regression, there has been widespread interest in cases where the penalty function corresponds to a normal scale mixture.  Many estimators in this class share the favorable sparsity-inducing property (i.e.~heavy tails) of the Bayesian bridge model.  This includes the relevance vector machine of \citet{tipping:2001}; the normal/Jeffreys model of \citet{figueiredo:2003} and \citet{bae:mallick:2004}; the normal/exponential-gamma model of \citet{griffin:brown:2005}; the normal/gamma and normal/inverse-Gaussian \citep{caron:doucet:2008, griffin:brown:2010}; the horseshoe prior of \citet{Carvalho:Polson:Scott:2008a}; and the double-Pareto model of \citet{dunson:armagan:lee:2010}.

In virtually all of these models, the primary difficulty is the mixing rate of the MCMC used to sample from the joint posterior for $\bbeta$.  Most MCMC approaches in this realm use latent variables to make sampling convenient.  But this can lead to poor mixing rates, especially in cases where the fraction of ``missing information''---that is, the information in the conditional distribution for $\bbeta$ introduced by the latent variables---is large.  Section 3.3 of the paper by \citet{hans:2008} contains an informative discussion of this point.  We have also included an online supplement to the manuscript that extensively documents the poor mixing behavior of Gibbs samplers within this realm.

\begin{figure}[t]
\begin{center}
\includegraphics[width=5.5in]{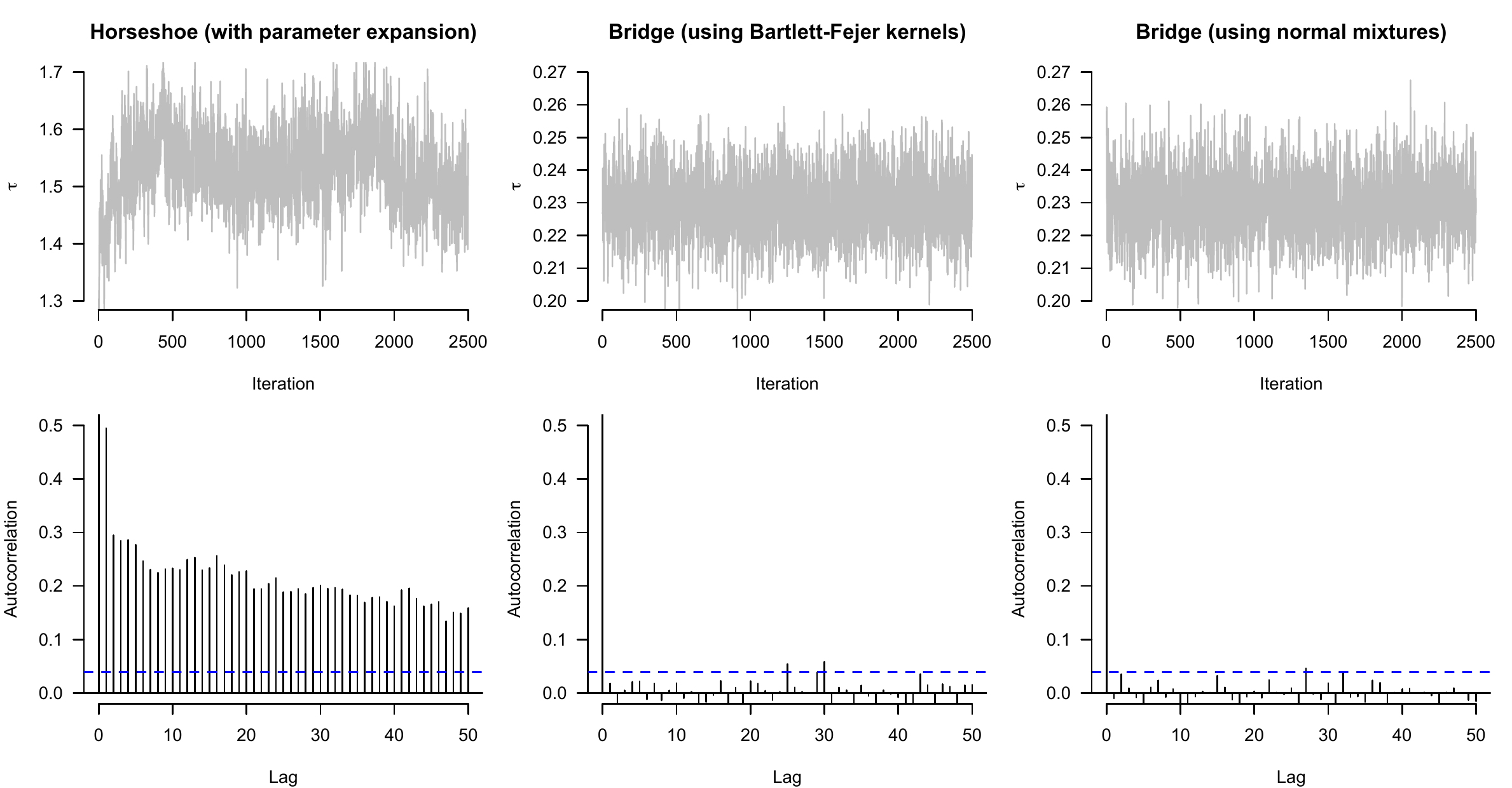} \\
\caption{\label{fig:bridgevshorseshoe} Comparison of the simulation histories for $\tau$, the global scale parameter, using MCMC for the bridge and the horseshoe on a 1000-dimensional orthogonal regression problem with $n=1100$ observations.  There were 100 non-zero entries in $\bbeta$ simulated from a $t_4$ distribution, and 900 zeros.  Because the priors have different functional forms, the $\tau$ parameters in each model have a comparable role but not a comparable scale, which accounts for the difference between the vertical axes.}
\end{center}
\end{figure}

In light of these difficulties, it comes as something of a surprise that the Bayesian bridge model leads to an MCMC strategy with an excellent mixing rate.  There are actually two such approaches, both of which exhibit very favorable mixing over $\tau$, the global scale parameter, which is known to be a serious problem in such models.  For example, Figure \ref{fig:bridgevshorseshoe} compares the performance of our bridge MCMC versus the best known Gibbs sampler for fitting the horseshoe prior \citep{Carvalho:Polson:Scott:2008a} on a 1000-variable orthogonal regression problem with 900 zero entries in $\bbeta$.  (See the supplement for details.)  The plots show the first 2500 iterations of the sampler, starting from $\tau=1$.  There is a dramatic difference in the effective sampling rate for $\tau$, which controls the overall level of sparsity in the estimate of $\bbeta$.  (Though these results are not shown here, equally striking differences emerge when comparing the simulation histories of the local scale parameters under each method.)
 
\subsection{Computational approach}

We would argue that the Bayesian bridge model is an interesting object for study on the basis of all three of these comparisons.  It leads to richer model summaries, superior performance in estimation and prediction, and better uncertainty quantification compared to the classical bridge.  It is better at handling sparsity than priors with exponential (or thinner) tails.  And it leads to an MCMC with superior mixing compared to other heavy-tailed, sparsity-inducing priors widely used in Bayesian inference.

These advantages, however, do not come for free.  In particular, posterior inference for the Bayesian bridge is more challenging than in most other Bayesian models of this type, where MCMC sampling relies upon representing the implied prior distribution for $\beta_j$ as a scale mixture of normals.   The exponential-power prior in (\ref{eqn:bridgeprior}) is known to lie within the normal-scale mixture class \citep{west:1987}.  Yet the mixing distribution that arises in the conditional posterior is that of an exponentially tilted alpha-stable random variable.  This complicates matters, due to the lack of a closed-form expression for the density function.  This fact was recognized by \citet{armagan:2009}, who proposed using variational methods to perform approximate Bayesian inference.

These issues can be overcome in two ways.  We outline our computational strategy here, and provide further details in Sections \ref{sec:DAstrategies}, \ref{sec:connections}, and \ref{sec:MCMC}.  The R package \verb|BayesBridge|, freely available online, implements all methods and experiments described in this paper.

\begin{figure}[t]
\begin{center}
\includegraphics[width=5in]{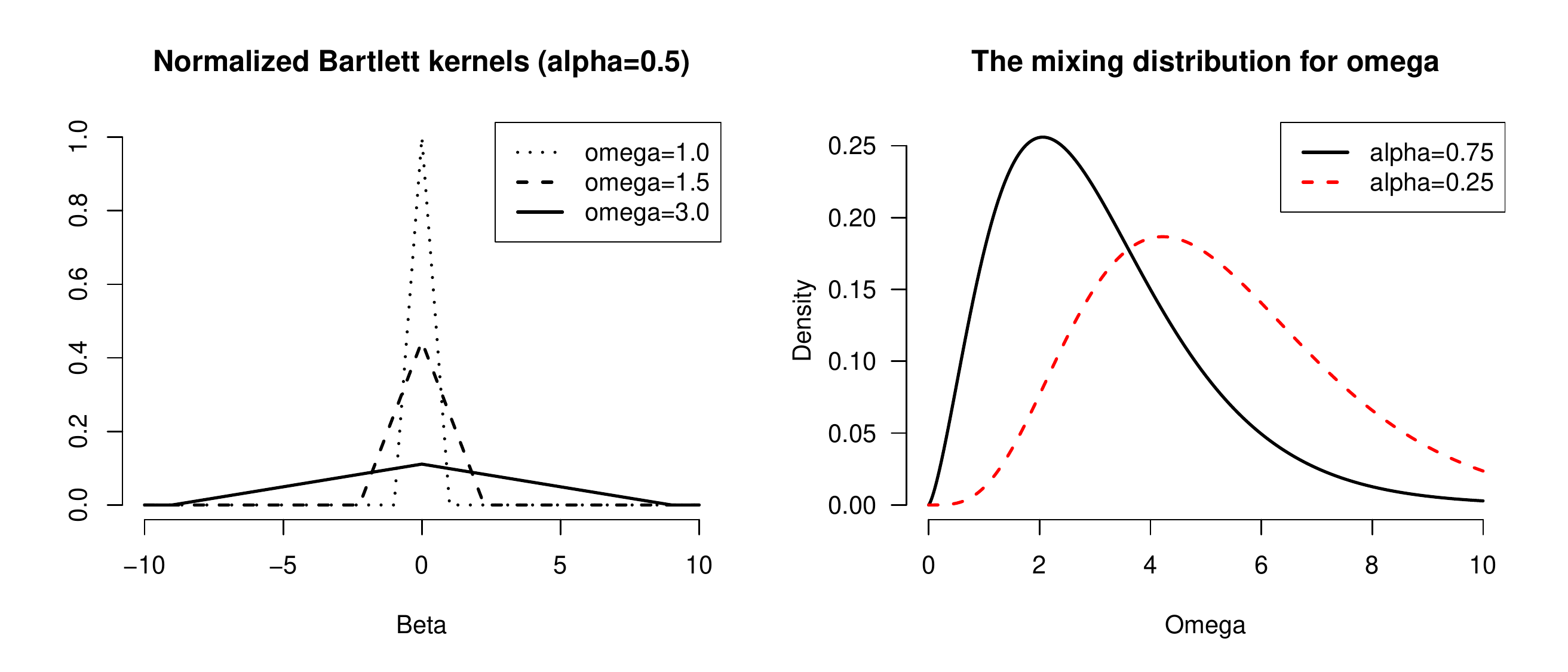}
\caption{\label{fig:bartlettmixtures} Left: triangular densities, or normalized Bartlett--Fejer kernels, of different widths.  Right: two examples of mixing distributions for $\omega_j$ that give rise to exponential-power marginals for $\beta_j$ in conjunction with the Bartlett--Fejer kernel.}
\end{center}
\end{figure}

The first approach is to work directly with normal mixtures of stable distributions, using rejection sampling or some other all-purpose algorithm within the context of a Gibbs sampler.  Some early proposals for sampling stable distributions can be found in \citet{devroye:1996} and \citet{godsill:2000}.  Neither of these proved to be sufficiently robust in our early implementations of the method.  But a referee pointed us to a much more recent algorithm from \citet{devroye:2009}.  The method is somewhat complicated, but seems very robust, and leads to generally excellent performance (see the empirical results in the online supplement).  Given current technology, it appears to be the best method for sampling the bridge model when the design matrix exhibits strong collinearity.

There is also a second, novel approach that turns out to be more efficient than the mixture-of-normals MCMC when the design matrix is orthogonal, or nearly so.  Specifically, we appeal to the following mixture representation, which is a special case of a more general result based on the Schoenberg--Williamson theorem for $n$-monotone densities:
\begin{eqnarray}
(y \mid \bbeta, \sigma^2) &\sim& \N(X\bbeta, \sigma^2 I) \nonumber \\
p(\beta_j \mid \tau, \omega_j, \alpha) &=& \frac{1}{\tau \omega_j^{1/\alpha}} \cdot \left \{ 1 - \left| \frac{\beta_j}{\tau \omega_j^{1/\alpha}} \right| \right \}_+ \label{eqn:bridgeBM1} \\
(\omega_j \mid \alpha) &\sim& \frac{1+\alpha}{2} \cdot \Ga(2+1/\alpha,1) + \frac{1-\alpha}{2} \cdot \Ga(1+1/\alpha,1)  \label{eqn:bridgeBM2} \, .
\end{eqnarray}
This scale mixture of triangles, or Bartlett-Fejer kernels, recovers $e^{-Q_{\by}(\bbeta)}$ as the marginal posterior in $\bbeta$.  The mixing distribution is depicted in Figure \ref{fig:bartlettmixtures} and explained in detail in Section \ref{sec:bartlettmixtures}.  It leads to a simple MCMC that avoids the need to deal with alpha-stable distributions, and can easily hop between distinct modes in the joint posterior.  (This is aided by the fact that the mixing distribution for the local scale $\omega_j$ has two distinct components.)

This alleviates one of the major outstanding difficulties of working with the bridge objective function and leads to noticeable gains in the orthogonal case.  As Section \ref{sec:DAstrategies} will show, it is also of interest in its own right, and can be used to represent many other penalty functions and likelihoods that arise in high-dimensional inference.  Our main theorem leads to an explicit Bayesian representation for any non-convex penalty function whose corresponding density version is proper.  This is a very wide class of penalties that can be accommodated via data augmentation.

\section{Data augmentation for the bridge model}

\label{sec:DAstrategies}

\subsection{As a scale mixture of normals}

We begin by discussing the two different data augmentation strategies that facilitate posterior inference for the Bayesian bridge model.

First, there is the mixture-of-normals representation, well known since \citet{west:1987}.  This can be seen by appealing to Bernstein's theorem, which holds that a function $f(x)$ is completely monotone if and only if it can be represented as a Laplace transform of some distribution function $G(\lambda)$:
\begin{equation}
\label{eqn:bernsteinmixture}
f(x) = \int_0^{\infty} e^{-s x} d G(s) \, .
\end{equation}
To represent the exponential-power prior as a Gaussian mixture for $\alpha \in (0,2]$, let $x = t^2/2$.  We then have
\begin{equation}
\label{eqn:eppmixture}
\exp(-|t|^{\alpha}) = \int_0^{\infty} e^{-s t^2/2} g (s) \ d s\, ,
\end{equation}
where $g(s)$ can be identified by recognizing the left-hand side as the Laplace transform, evaluated at $t^2/2$, of a positive alpha-stable random variable with index of stability $\alpha/2$ \citep[also see][]{Polson:Scott:2010b}.

Similar Gaussian representations have been exploited to yield conditionally conjugate MCMC algorithms for a variety of models, such as the lasso and the horseshoe priors.  Unfortunately, the case of the bridge is less simple.  To see this, consider the joint posterior implied by (\ref{eqn:bridgeobjective}) and (\ref{eqn:eppmixture}):
\begin{align}
p( \bbeta , \Lambda \mid y) & = C \exp \left ( - \nu^{ 2/\alpha } \bbeta^\prime \Lambda \bbeta
 - \frac{1}{2 \sigma^2} \bbeta^\prime X^\prime X \bbeta  + \bbeta^\prime \sigma^{-2} X^\prime \by \right )
 \prod_{j=1}^p p ( \lambda_j ) \nonumber \\
 & = C \exp \left \{ - \frac{1}{2} \bbeta^\prime \left ( \sigma^{-2} X^\prime X 
 + 2 \nu^{ { 2/\alpha }  } \Lambda \right ) \bbeta 
 + \bbeta^\prime \sigma^{-2} X^\prime \by \right \} \prod_{j=1}^p p (\lambda_j) \label{eqn:normaljoint1} \, , 
\end{align}
where $\Lambda = \mbox{diag}(\lambda_1, \ldots, \lambda_j)$, and $p(\lambda_j) = \lambda_j^{-1/2} g(\lambda_j) $, $g$ denoting the stable density from the integrand in (\ref{eqn:eppmixture}).  The conditional posterior of $ \lambda_j $
given $ \beta_j $ is then an exponentially tilted stable random variable,
$$
p( \lambda_j \mid \beta_j ) = \frac{ e^{ - \nu^{\frac{2}{\alpha}} | \beta_j |^2 \lambda_j } p (\lambda_j) }{ 
 \mathbb{E} \left ( e^{- \nu^{\frac{2}{\alpha}} | \beta_j |^2 \lambda_j } \right ) } \, ,
$$
with the expectation in the denominator taken over the prior.  Neither the prior nor posterior for $\lambda_j$ are known in closed form, and can be only be written explicitly as an infinite series.

\subsection{An alternative approach for $n$-monotone densities}
\label{sec:bartlettmixtures}

Bernstein's theorem holds for completely monotone density functions, and can be used to construct scale mixtures of normals by evaluating the right-hand side of (\ref{eqn:bernsteinmixture}) at $t^2/2$.  As we have seen in the case of the bridge, this results in a conditionally Gaussian form for the parameter of interest, but a potentially difficult mixing distribution for the latent variable.

We now construct an alternate data-augmentation scheme that avoids these difficulties.  Specifically, consider the class of symmetric density functions $f(x)$ that are $n$-monotone on $(0, \infty)$ for some integer $n$: that is, $(-1)^k f^{(k)} (|x|) \geq 0$ for $k = 0, \ldots, n-1$, where $f^{(k)}$ is the $k$th derivative of $f$, and $f^{(0)} \equiv f$.    

The following result builds on a classic theorem of Schoenberg and Williamson.  It establishes that any $n$-monotone density $f(x)$ may be represented as a scale mixture of betas, and that we may invert for the mixing distribution using the derivatives of $f$.

\begin{theorem} 
\label{thm:kmonotone}
Let $f(x)$ be a bounded density function that is symmetric about zero and $n$-monotone over $(0, \infty)$, normalized so that $f(0) = 1$.  Let $C = \{2\int_0^{\infty} f(t) \ dt\}^{-1}$ denote the normalizing constant that makes $f(x)$ a proper density on the real line.  Then $f$ can be represented as the following mixture for any integer $k$, $1 \leq k \leq n$:
\begin{equation}
\label{eqn:kmonomixture}
C f(x) =  \int_0^\infty 
\frac{1}{s} \  k  \left( 1 -  \frac{|x|}{s} \right)^{k-1}_+ \ g(s)  ds \, ,
\end{equation}
where $a_{+} = \max(a,0)$, and where the mixing density $g(s)$ is
$$
g(s) = C k^{-1}  \sum_{j=0}^{k-1} \frac{(-1)^j }{j!} \left\{ j s^{j} f^{(j)}(s)   +  s^{j+1} f^{(j+1)}(s) \right\} \, .
$$
\end{theorem}
Crucially, the mixing density in the $k$-monotone case has only a finite number of terms.  Moreover, a function that is completely monotone is also $n$-monotone for all finite $n$.  Thus the proposition applies to any function for which Bernstein's theorem holds, allowing an arbitrary (presumably convenient) choice of $n$.

Return now to the Bayesian bridge model.  The exponential power density for $0 < \alpha \leq 1$ is completely monotone on the positive reals, and therefore any value of $k$ may be used in Equation (\ref{eqn:kmonomixture}).   We focus on the choice $k=2$.  The kernel functions that arise here have been referred to as Bartlett kernels in econometrics, a usage which appears to originate in a series of papers by Newey and West on robust estimation.  They have also been called Fejer densities in probability theory; see \citet{dugue:girault:1955}, who study them in connection with the theory of characteristic functions of Polya type.  Thus we refer to them as Bartlett--Fejer kernels.

\begin{corollary} 
\label{thm:bartlettmixture}
Let $f(x)$ be a function that is symmetric about the origin; integrable, convex, and twice-differentiable on $(0, \infty)$; and for which $f(0) = 1$.  Let $C = \{2\int_0^{\infty} f(t) \ dt\}^{-1}$ denote the normalizing constant that makes $f(x)$ a density on the real line.  Then $f$ is the following mixture of Bartlett--Fejer kernels:
\begin{equation}
C f(x) =  \int_0^\infty 
\frac{1}{s} \left \{ 1 - \frac{|t|}{s} \right \}_+ C s^2 f''(s) \ ds \, ,
\end{equation}
where $a_{+} = \max(a,0)$.
\end{corollary}

The proof involves only simple manipulations, and is omitted.  Using this corollary, the exponential power density with $\alpha \in (0,1]$ can be represented in a particularly simple way.  To see this, transform $s \to \omega \equiv s^{\alpha}$ and observe that:
\begin{eqnarray*}
\frac{1}{2\tau} \exp(-|\beta/\tau|^{\alpha} ) &=& \int_0^\infty \frac{1}{\tau}  \left \{ 1 - \left| \frac{\beta}{\tau \omega^{1/\alpha}} \right| \right \}_+ p(\omega \mid \alpha) \ d \omega \\
p(\omega \mid \alpha) &=& \alpha \omega e^{-\omega} + (1-\alpha) e^{-\omega} \, .
\end{eqnarray*}
Simple algebra with the normalizing constants yields a properly normalized mixture of Bartlett--Fejer kernels:
\begin{eqnarray*}
\frac{\alpha}{2\tau \Gamma(1 + 1/\alpha)} \exp(-|\beta/\tau|^{\alpha}| ) &=& \int_0^\infty \frac{1}{\tau \omega^{1/\alpha} }  \left \{ 1 - \left| \frac{\beta}{\tau \omega^{1/\alpha}} \right| \right \}_+ p(\omega \mid \alpha) \ d \omega \\
p(\omega \mid \alpha) &=& \frac{1+\alpha}{2} c_1 \omega^{1 + 1/\alpha} e^{-\omega} + \frac{1-\alpha}{2} c_2 \omega^{1/\alpha} e^{-\omega} \, ,
\end{eqnarray*}
This is a simple two-component mixture of gammas, where $c_1$ and $c_2$ are the normalizing constants of each component.  The Bayesian lasso is a special case, for which the second mixture component drops out.

\section{The connection with other latent-variable methods}

\label{sec:connections}

The latent-variable scheme suggested by Theorem \ref{thm:kmonotone} was originally motivated by the potential inefficiencies of working with exponentially tilted stable random variables, and does lead to noticeable improvements in the orthogonal case.  But the representation is of considerable interest in its own right, quite apart from its application to the bridge model.

The analogy with slice sampling is instructive.  In both cases, the basic problem is to sample from a posterior distribution of the form
$L(\theta) p(\theta)/Z$, where $L$ is a likelihood, $p$ is a prior, and $Z$ is the normalization constant. For example, if we slice out the prior, we introduce an auxiliary variable $u$, conditionally uniform on  $0\leq u < p(\theta ) $, and sample from the joint distribution
$$
\pi ( \theta ,u) = \mathbb{I} \{ u < p(\theta ) \} L(\theta )/Z \, ,
$$
where $\mathbb{I}(\cdot)$ is the indicator function.   The posterior of interest is simply the marginal distribution for $\theta$.

The difficulty is that, given $u$, one needs to be able to calculate the slice region where $p(\theta ) > u$.  In our data-augmentation approach, the analogous inversion problem is already done: it reduces to the set where $ |\theta |<\omega $.  For example, in the once-monotone case, we have
$$
\pi ( \theta , \omega ) = \mathbb{I} \left ( | \theta | < \omega \right ) g(\omega) L(\theta )/Z \, ,
$$
where $\omega$ now plays a role similar to that of $u$.  We have removed the problem of inverting a slice region, at the cost of generating $\omega$ from a non-uniform distribution $g(\omega)$, which is uniquely identified by Theorem \ref{thm:kmonotone}.  Of course, the question of which method leads to simpler calculations will depend on context.  We simply point out that there are many cases where inverting a slice region is nontrivial.  See \citet{damien:etal:1999}, \citet{roberts:rosenthal:2002}, or \citet{neal:2003}.

Moreover, the once-monotone case is just one example of the wider family.  It is well known that the efficiency of a latent-variable scheme is inversely related to the amount of information about $\theta$ conveyed by the latent variable.  The clear lesson is that, all else being equal, one should mimic the target density as closely as possible when choosing a kernel over which to mix.  Theorem 2.1 gives the designer of sampling algorithms wide latitude in this regard.  For example, in the twice-monotone case, we have
$$
\pi ( \theta , \omega ) = \mathbb{I} \left ( | \theta | < \omega \right ) g(\omega) 
\left (1 - |\theta|/\omega \right ) L(\theta )/Z \, .
$$
For the Bayesian bridge model, this representation is especially attractive.  The exponential-power density simply looks more like a triangle than like a rectangle.  Additionally, the bimodality of each marginal posterior for $\beta_j$ is nicely matched by the fact that $g(\omega_j)$ has two distinct components.  (See Figure \ref{fig:multimodal}.)

Moving beyond the once-monotone does pose its own trade-offs.  The mixing distribution $g(\omega)$ potentially becomes more complicated, and we must now sample from the tilted distribution with density proportional to $ \left (1 - |\theta|/\omega \right )^{k-1} L(\theta ) $.  The difficulty of this step will also depend on context.  In the twice-monotone representation of the bridge model, it turns out to be straightforward.  In other cases, one may appeal to the algorithm of \citet{stein:keblis:2009} for simulating the triangle distribution, which can be extended to the case of a triangle times another density.  Other general strategies for sampling tilted densities are alluded to in \citet{devroye:2009}.

Finally, there is an important connection between our result and the Gaussian mixture representation, which parallels the relationship between the Schoenberg--Williamson and Bernstein theorems.  To see this, let $u = k/s$.  Observe that we obtain the completely monotonic case as $k$ diverges:
\begin{eqnarray*}
f(x) &\propto&  \int_0^\infty 
 \   \left( 1 -  \frac{u x}{k} \right)^{k-1}_+ \ d P(u) \\
 &\to & \int_0^\infty 
 \   e^{-sx} d \tilde{P}(s) \
\end{eqnarray*}
for positive $x$ and a suitably defined limiting measure $\tilde{P}(s)$, into which a factor of $s$ has been implicitly absorbed.  By evaluating this at $s=t^2/2$, we obtain a scale mixture of normals as a limiting case of a scale mixture of betas.  The inversion formula, too, is similar.  In particular, for the case of the exponential power kernel, we have
$$
\exp \left ( - | x |^\alpha \right ) = \int_0^\infty e^{- x s} g(s) ds \quad {\rm with} \quad 
g(s) = \sum_{j=1}^\infty (-1)^j \frac{ s^{- j \alpha -1 }}{ j! \Gamma ( - \alpha j )} \, ,
$$
which clearly parallels the expression given in Theorem \ref{thm:kmonotone}.

Although a long discussion here would lead us astray from our main point, there may be many interesting cases where the new approach could prove fruitful.  One such example is the type-I extreme value distribution, $p(x) = \exp(-x - e^{-x})$.  From Theorem \ref{thm:bartlettmixture}, we have
$$
e^{-x} = \int_0^{\infty} \left(1 - \frac{|x|}{\omega} \right)_+ e^{-\omega} d \omega \, ,
$$
and therefore $e^{-e^{-x}}$ can be written in terms of a gamma mixing measure:
$$
e^{-e^{-x}} = \int_0^{\infty} \frac{1}{\omega} \left(1 - \frac{e^{-x}}{\omega} \right)_+ \omega e^{-\omega} d \omega \, .
$$

\section{MCMC sampling for the Bayesian bridge}

\label{sec:MCMC}

\subsection{Overview of approach}

For sampling the Bayesian bridge posterior, we recommend a hybrid computational approach, which we have implemented as the default setting in our \verb|BayesBridge| R package.  Due to space constraints, the evidence supporting this recommendation is detailed in an online supplemental file, where we describe the results of an extensive benchmarking study.  We briefly summarize our conclusions here.

When the design matrix $X$ exhibits strong collinearity, the normal scale mixture representation is the better choice.  In data sets involving many higher-order interaction terms, the efficiency advantage can be substantial.  On the other hand, when the design matrix is orthogonal, the Bartlett-Fejer representation usually leads to an effective sampling rate roughly two to three times that of the Gaussian method.  Orthogonal designs arise in principal-components regression, and in many nonlinear problems where the effect of a covariate is expanded in an orthogonal basis.  They also arise when using the generalized $g$-priors for $p>n$ problems discussed in \citet{Polson:Scott:2010b}.  Finally, the representation is relevant to cases where one wishes to model a system with an exponential-power likelihood, where conditional indepedence of each observation will mimic the orthogonal case in regression analysis.  See, for example, \citet{godsill:2000} for an application in acoustics.

We use the method described in  \citet{devroye:2009} for sampling exponentially tilted alpha-stable random variables.  With this capability in place, it is easy to use (\ref{eqn:normaljoint1}) to generate posterior draws, appealing to standard multivariate normal theory.  Thus we omit a long discussion of the normal-mixture method, and focus on the mixture-of-betas approach.

\subsection{Sampling $\bbeta$ and the latent variables}

To see why the representation in (\ref{eqn:bridgeBM1})--(\ref{eqn:bridgeBM2}) leads to a simple algorithm for posterior sampling, consider the joint distribution for $\bbeta$ and the latent $\omega_j$'s:
\begin{equation}
\label{eqn:bridgewithlatents}
p( \bbeta , \Omega  \mid \tau , y ) = C \exp \left ( - \frac{1}{2 \sigma^2} \bbeta^\prime X^\prime X \bbeta 
 + \frac{1}{\sigma^2} \bbeta^\prime X^\prime y \right ) \prod_{i=1}^p p( \omega_j \mid \alpha )
 \prod_{i=1}^p \left ( 1 - \frac{ |\beta_j| }{\tau \omega_j^{1/\alpha}} \right )_+ \, .
\end{equation}
Introduce further slice variables $u_1, \ldots, u_j$.  This leads to the joint posterior
\begin{eqnarray}
p( \bbeta , \Omega , u \mid \tau , \by ) &\propto& \exp \left ( - \frac{1}{2 \sigma^2} \bbeta^\prime X^\prime X \bbeta 
 + \frac{1}{\sigma^2} \bbeta^\prime X^\prime y \right ) \nonumber \\
 &\times& \prod_{j=1}^p p( \omega_j \mid \alpha )
 \prod_{j=1}^p \mathbb{I} \left ( 0 \leq u_j \leq \left[ 1 - \frac{ |\beta_j| }{\tau \omega_j^{1/\alpha}} \right] \right )\label{eqn:bridgeslicevars} \, .
\end{eqnarray}
Note that we have implicitly absorbed a factor of $\omega^{1/\alpha}$ from the normalization constant for the Bartlett--Fejer kernel into the gamma conditional for $\omega_j$. 

Applying Corollary \ref{thm:bartlettmixture}, if we marginalize out both the slice variables and the latent $\omega_j$'s, we recover the Bayesian bridge posterior distribution,
$$
p( \bbeta \mid \by) = C \exp \left ( - \frac{1}{2 \sigma^2} \Vert \by- X \bbeta \Vert^2  - \sum_{j=1}^p 
| \beta_j /\tau |^\alpha  \right ) \, .
$$

We can invert the slice region in (\ref{eqn:bridgeslicevars}) by defining $(a_j,b_j)$ as
$$
| \beta_j | \leq \tau^{-1} ( 1- u_j )\omega^{1/\alpha }_j = b_j \; \; {\rm and} \; \;
 \omega_j \geq \left ( \frac{ | \beta_j / \tau | }{ 1- u_j} \right )^\alpha = a_j \, .
$$
This leads us to an exact Gibbs sampler that starts at initial guesses for $ ( \bbeta, \Omega ) $ and iterates the following steps:
\begin{enumerate}
\item Generate $ (u_j \mid \beta_j ,\omega_j) \sim \mbox{Unif} \left ( 0 , 1 -  |\beta_j / \tau| \omega_j^{- 1/\alpha} \right ) $ .
\item Generate each $ \omega_j $ from a mixture of truncated gammas, as described below.
\item Update $ \bbeta$ from a truncated multivariate normal proportional to
$$
\N \left ( \hat{\bbeta} , \sigma^2 ( X^\prime X )^{-1} \right ) 
\mathbb{I} \left ( | \beta_j | \leq b_j \; \mbox{for all} \; j \right ) \, ,
$$
where $\hat{\bbeta}$ indicates the least-squares estimate for $\bbeta$.
\end{enumerate}

We explored several different methods for simulating from the truncated multivariate normal, ultimately settling on the proposal of \citet{rodriguez-yam-etal-2004} as the most efficient.  (It is important to observe that, using this method, $\bbeta$ cannot be regenerated at each step, merely updated component-wise.) The conditional posterior of the latent $\omega_j$'s can be determined as follows.  Suppressing subscripts for the moment, we may write the compute conditional for $\omega$ as:
\begin{eqnarray*}
p ( \omega \mid \alpha ) & = &\alpha ( \omega e^{- \omega} ) + ( 1- \alpha ) e^{- \omega}\\
p ( \omega \mid a , \alpha ) & =& C_a \left \{ \alpha ( \omega e^{- \omega} ) + ( 1- \alpha ) e^{- \omega}
 \right \} \mathbb{I} \left ( \omega \geq a \right ) \, ,
\end{eqnarray*}
where $a$ comes from inverting the slice region in (\ref{eqn:bridgeslicevars}) and $ C_a$ is the normalization constant.  

We can simulate from this mixture of truncated gammas by defining $\bar{\omega} =  \omega - a$, where $\bar{\omega} > 0$.  Then $ \bar{\omega} $ has density
\begin{align*}
p( \bar{\omega}| a , \alpha ) & = C_a \left \{ \alpha  e^{-a} ( a + \bar{\omega} ) e^{- \bar{\omega}}
 + (1 - \alpha ) e^{-a} e^{- \bar{\omega} } \right \} \\
 & =  \frac{ \alpha}{1 + \alpha a} \cdot \bar{\omega} e^{- \bar{\omega}}
 + \left\{ \frac{ 1 - \alpha (1-a)}{1 + \alpha a} \right\} \cdot e^{- \bar{\omega}} \, .
\end{align*}
This is a mixture of gammas, where
$$
(\bar{\omega} \mid a) \sim
\left\{
\begin{array}{c c}
\Gamma (1,1) \quad  {\rm with \; prob} \;& \frac{ 1 - \alpha (1-a) }{1 + \alpha a} \\   
\Gamma (2,1) \quad  {\rm with \; prob} \;& \frac{ \alpha  }{1 + \alpha a} \, .
\end{array}
\right. 
$$
After sampling $\bar{\omega}$, simply transform back using the fact that $\omega = a + \bar{\omega}$.

This representation has two interesting and intuitive features.  First, full conditional for $\bbeta$ in step 3 is centered at the usual least-squares estimate $\hat{\bbeta}$.  Only the truncations ($b_j$) change at each step, which eliminates matrix operations.

\begin{figure}[t]
\begin{center}
\includegraphics[width=6in]{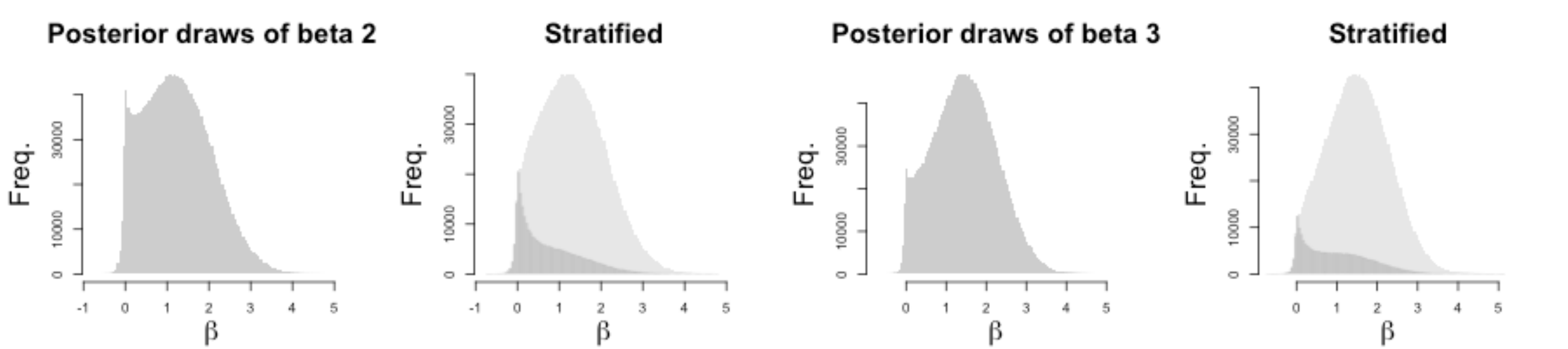}
\caption{\label{fig:multimodal} Left two panes: posterior draws for $\beta_2$ in the simulated example  The far-left plot shows the marginal draws, while the plot second from left shows the draws stratified by the mixture component of the most recent draw for $\omega_2$.  Right two panes: the same two plots are repeated for $\beta_3$, which in this case facilitate the identification of both modes.}
\end{center}
\end{figure}

Second, the mixture-of-gammas form of $p(\omega)$ naturally accounts for the bimodality in the marginal posterior distribution, $p( \beta_j \mid \by ) = \int p( \beta_j \mid \omega , \by ) p ( \omega_j \mid \by ) d \omega_j $.  In some cases, in fact, each mixture component of the conditional for $\omega_j$ represents a distinct mode of the marginal posterior for $\beta_j$.  As Figure \ref{fig:multimodal} shows, this endows the algorithm with the ability to explore the multiple modes of the joint posterior.  These plots come from a simulated data set with $p=20$ and $n=200$, with the predictors having pairwise positive correlation of $0.99$.  We set $\tau = 0.1$ and $\alpha = 0.85$ and ran the mixture-of-triangles MCMC for 200,000 iterations.  (These settings were arbitrary, but identified the effect most clearly out of the several we tried.)  The figure shows the posterior draws for two of the coefficients ($\beta_2$ and $\beta_3$), stratified by the mixture component for the most recent draw for the corresponding $\omega_j$.  In the case of $\beta_3$, the stratification helps to identify a mode which might easily be missed in a histogram of draws from the marginal posterior.

\subsection{Sampling hyperparameters}

To update the global scale parameter $\tau$, we work directly with the exponential-power density, marginalizing out the latent variables $\{\omega_j, u_j\}$.  This is a crucial source of efficiency in the bridge MCMC, and leads to the favorable mixing evident in Figure \ref{fig:bridgevshorseshoe}.  From (\ref{eqn:bridgeobjective}), observe that the posterior for $\nu \equiv \tau^{-\alpha}$, given $\bbeta$, is conditionally independent of $\by$, and takes the form
$$
p(\nu \mid \bbeta) \propto \nu^{p/\alpha} \exp(-\nu \sum_{j=1}^p |\beta_j|^{\alpha}) \ p(\nu) \, .
$$
Therefore if $\nu$ has a Gamma($c,d$) prior, its conditional posterior will also be a gamma distribution, with hyperparameters $c^{\star} = c + p/\alpha$ and $d^{\star} = d + \sum_{j=1}^p |\beta_j|^{\alpha}$.  To sample $\tau$, simply draw $\nu$ from this gamma distribution, and use the transformation $\tau = \nu^{-1/\alpha}$.  Alternative priors for $\nu$ can also be considered, in which case the gamma form of the conditional likelihood in $\nu$ will make for a useful proposal distribution that closely approximates the posterior.  As Figure \ref{fig:bridgevshorseshoe} from the introduction shows, the ability to marginalize over the local scales in sampling $\tau$ is crucial here in leading to a good mixing rate.

\begin{figure}[t]
\begin{center}
\includegraphics[width=4in]{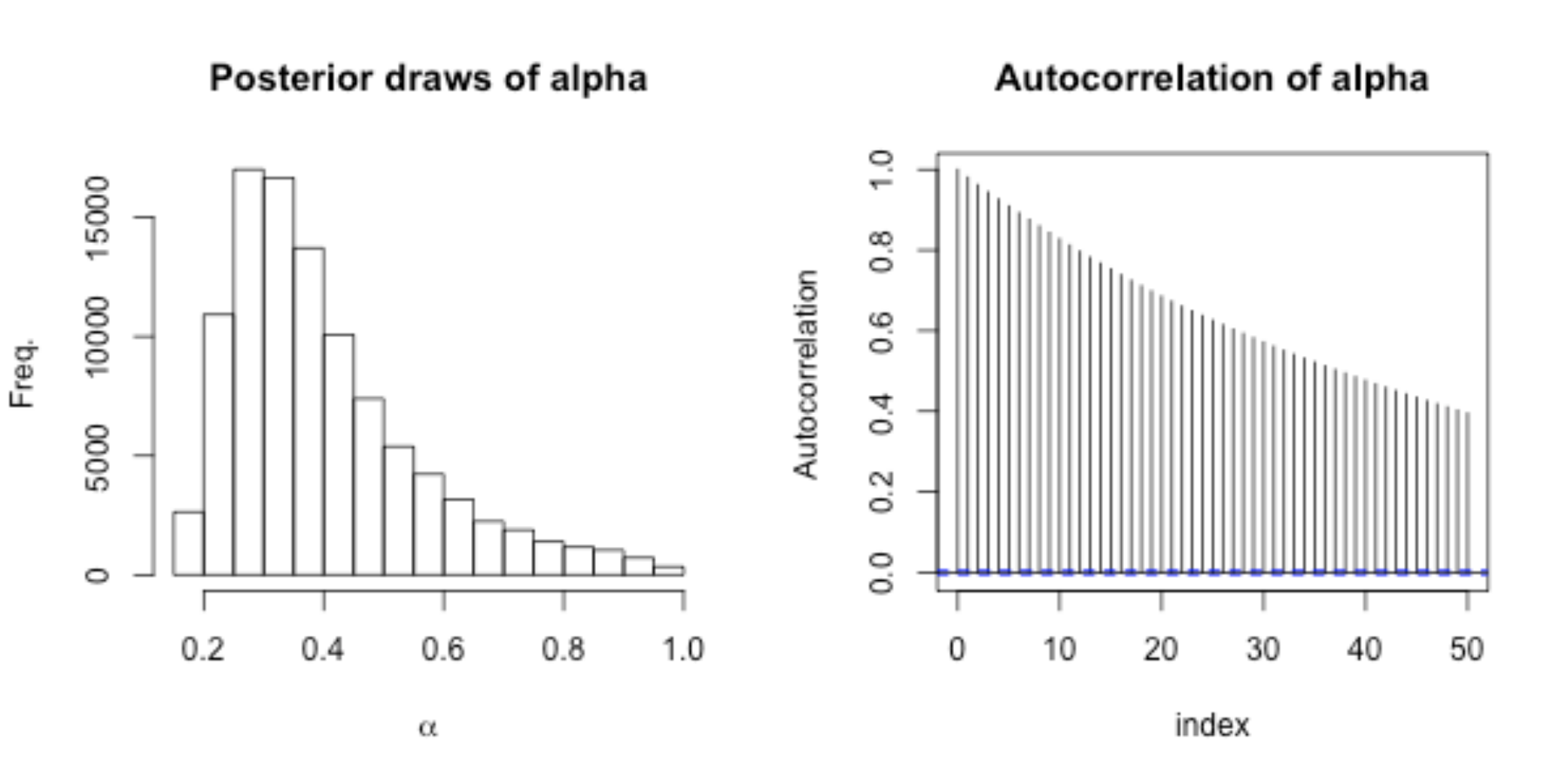}
\caption{\label{fig:alpha-db} Left: The posterior distribution of the concavity parameter $\alpha$ for the diabetes data, using a random-walk Metropolis sampler.  Right: the autocorrelation function of the posterior draws.}
\end{center}
\end{figure}

In many cases the concavity parameter $\alpha$ will be fixed ahead of time to reflect a particular desired shape of the penalty function.  But it too can be give a prior $p(\alpha)$, most conveniently from the beta family, and can be updated using a random-walk Metropolis sampler.  Figure \ref{fig:alpha-db} shows the results of using this sampler under a uniform prior for $\alpha$ in the well-known data set on blood glucose levels in diabetes patients, available in the R package \verb|lars| \citep{efron:LARS:2004}.  There are only 10 predictors for this problem, and therefore a considerable amount of uncertainty about the value of $\alpha$.  It is interesting, however, how much the posterior is pulled away from $\alpha=1$, which corresponds to the lasso prior.

Similar MCMC algorithms can be used to fit Bayesian analogues of bridge-penalized logistic regression and quantile regression:
\begin{eqnarray*}
\hat{\bbeta}_{LR} &=& \arg \min_{\beta \in \mathbb{R}^p} \left\{  \sum_{i=1}^n \log(1 + \exp\{ -y_i x_i' \bbeta \} )
 + \sum_{j=1}^p |\beta_j / \tau |^{\alpha}  \right) \\
\hat{\bbeta}_{QR} &=& \arg \min_{\beta \in \mathbb{R}^p} \left\{  \sum_{i=1}^n\{ |y_i -x_i' \bbeta| + (2q-1) (y_i -x_i' \bbeta) \}
 + \sum_{j=1}^p |\beta_j / \tau |^{\alpha}  \right) \, ,
\end{eqnarray*}
where binary outcomes are encoded as $\pm 1$, or where $q \in (0,1)$ is the desired quantile.  To fit these estimators within a Bayesian framework, once introduces a second set of latent variables corresponding to the $n$ individual terms in the likelihood.  This allows the quantile-regression and logit likelihoods to be represented as mean--variance mixtures of Gaussian models with respect to known mixing measures.  The corresponding distributional theory is described in detail by \citet{Polson:Scott:2011a} and \cite{polson:scott:windle:2012a}.

\section{Examples}
\label{sec:examples}

\subsection{Diabetes data}

\begin{figure}
\begin{center}
\includegraphics[width=6in]{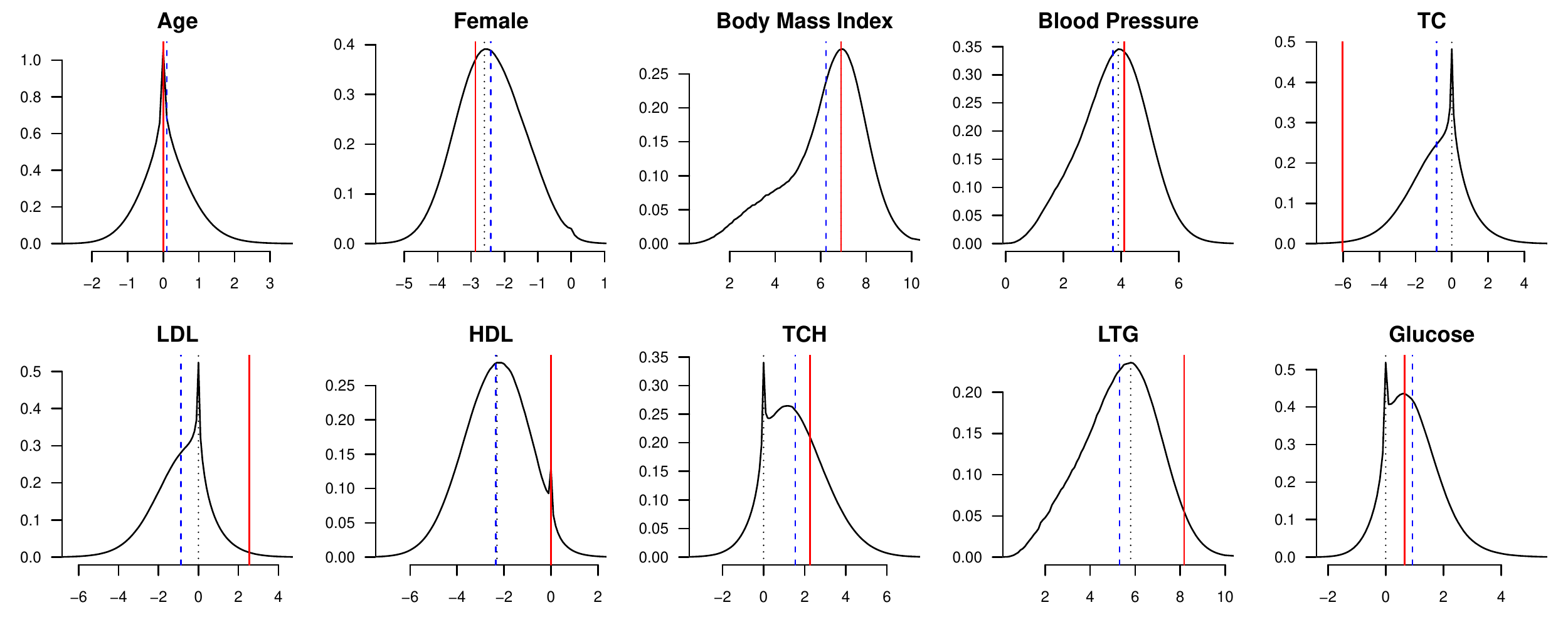}
\caption{\label{fig:diabetes-betas} Marginal posterior densities for the marginal effects of 10 predictors in the diabetes data.  Solid line: penalized-likelihood solution with $\nu$ chosen by generalized cross validation.  Dashed line: marginal posterior mean for $\beta_j$.  Dotted line: mode of the marginal distribution for $\beta_j$ under the fully Bayes posterior.  All predictors were standardized.}
\end{center}
\end{figure}

We first explore the Bayesian bridge estimator using the diabetes data, already described.  We fit the Bayesian bridge using a default Gamma(2,2) prior for $\nu$.  We also fit the classical bridge, using generalized cross validation and the EM algorithm from \citet{Polson:Scott:2011a}.  Both the predictor and responses were centered, while the predictors were also re-scaled to have unit variance.  At each step of the MCMC for the Bayesian model, we calculated the conditional posterior density for each $\beta_j$ at a discrete grid of values.   It is striking that, even for this relatively information-rich problem (10 predictors and 442 observations), significant differences emerge between the Bayesian and classical methods.

Figure \ref{fig:diabetes-betas} summarizes the results of the two fits, showing both the marginal posterior density and the classical bridge solution for each of the 10 regression coefficients.  One notable feature of the problem is the pronounced multimodality in the joint posterior distribution for the Bayesian bridge.  Observe, for example, the two distinct modes in the marginal posteriors for the coefficients associated with the TCH and Glucose predictors (and, to a lesser extent, for the HDL and Female predictors).  In none of these cases does it seem satisfactory to summarize information about $\beta_j$ using only a single number, as the classical solution forces one to do.

Second, observe that the classical bridge solution does not coincide with the joint mode of the fully Bayesian posterior distribution.  This discrepancy can be attributed to uncertainty in $\tau$ and $\sigma$, which is ignored in the classical solution.  Marginalizing over these hyperparameters leads to a fundamentally different objective function, and therefore a different joint posterior mode.

The difference between the classical mode and the Bayesian mode, moreover, need not be small.  Consider the posterior distributions for the TC and LDL coefficients.  These two predictors have a sample correlation of $-0.897$.  The Bayesian solution concentrates in a region of $\mathbb{R}^p$ where neither of these coefficients exerts much of an effect.  The classical solution, on the other hand, says that both predictors should be in the model with large coefficients of opposite sign.

It is impossible to say in any objective sense whether TC and HDL are both necessary, or instead are redundant copies of the same unhelpful information.  It is highly surprising, however, that such a marked difference would arise between the full Bayes mode and the classical mode, and that this difference would fundamentally alter one's conclusions about two predictors out of ten.  The full Bayes posterior mean is, of course, different yet again.  Clearly a very important role here is played by the decision of whether (and how) to account for uncertainty in $\tau$ and $\sigma$.

\subsection{Out-of-sample prediction results}

Next, we describe the results from three out-of-sample prediction exercises involving three benchmark data sets.  First, we used the Boston housing data: available in the R package \verb|mlbench|.  The goal is to predict the median house price for 506 census tracts of Boston from the 1970 census.  As covariates, we used the 14 original predictors, plus all interactions and squared terms for quantitative predictors.  Second, we used the data on ozone concentration, available in the R package \verb|mlbench|.  The goal is to predict the concentration of ozone in the atmosphere above Los Angeles using various environmental covariates.  As covariates, we used the 9 original predictors, plus all interactions and squared terms for quantitative predictors.  Finally, we used the NIR glucose data, available in the R package \verb|chemometrics|.   The goal is to predict the concentration of glucose in molecules using data from NIR spectroscopy.

For each data set, we created 100 different train/test splits, using the results from the training data to forecast the test data.  For each train/test split we estimated $\bbeta$ using least-squares, the classical bridge (using EM), and the Bayesian-bridge posterior mean using our MCMC method based on stable mixtures.  In all cases we centered and standardized the predictors, and centered the response.  For the classical bridge estimator, the regularization parameter $\nu$ was chosen by generalized cross validation; while for the Bayesian bridge, $\sigma$ was assigned Jeffreys' prior and $\nu$ a default Gamma(2,2) prior.  We used two different settings for $\alpha$ in the Bayesian bridge: one with $\alpha$ fixed at a default setting of $0.5$, and another with $\alpha$ estimated using a random-walk Metropolis step.  R scripts implementing all of these experiments are included as a supplemental file.

\begin{figure}
\begin{center}
\includegraphics[width=6in]{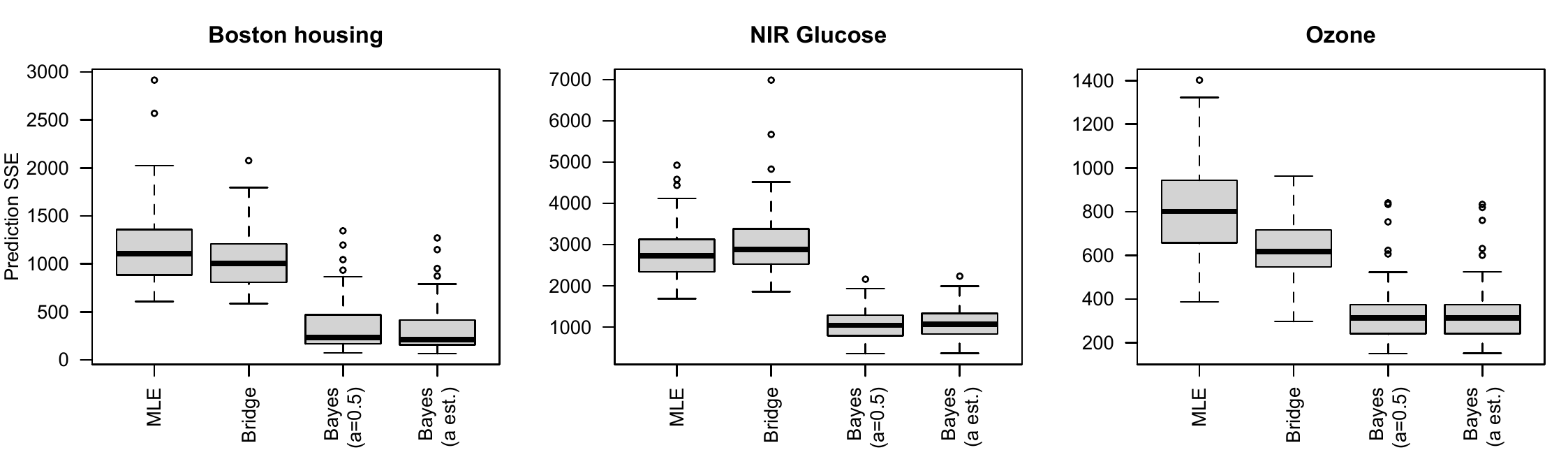}
\caption{\label{fig:predictionSSE} Boxplots of the sum of squared errors in prediction hold-out data using four different methods for estimating $\beta$.  From left to right, these are the MLE; the classical bridge with $\alpha=0.5$, and $\nu$ chosen by generalized cross validation; the Bayesian bridge, with $\alpha=0.5$; and the Bayesian bridge with $\alpha$ estimated under a uniform prior.}
\end{center}
\end{figure}

We measured performance of each method by computing the sum of squared errors in predicting $y$ on the test data set.  The results are in Figure \ref{fig:predictionSSE}.  In all three cases, the posterior mean estimator outperforms both least squares and the classical bridge estimator, and the fixed choice of $\alpha=0.5$ very nearly matched the performance of the model which marginalized over a uniform prior for $\alpha$.


\subsection{Simulated data with correlated design}

We conducted three experiments, all with $p=100$ and $n=101$, for $\alpha \in \{0.9, 0.7, 0.5\}$.  Each experiment involved 250 data sets constructed by: (1) simulating regression coefficients from the exponential power distribution for the given choice of $\alpha$; (2) simulating correlated design matrices $X$; and (3) simulating residuals from a Gaussian distribution.  In all cases we set $\sigma = \tau = 1$.  The rows of each design matrix were simulated from a Gaussian factor model, with covariance matrix $V = BB' + I$ for a $100 \times 10$ factor loadings matrix $B$ with independent standard normal entries.  As is typical for Gaussian factor models with many fewer factors (10) than ambient dimensions (100), this choice led to marked multi-collinearity among the columns of each simulated $X$.

For each simulated data set we again estimated $\bbeta$ using least squares, the classical bridge, and the Bayesian bridge posterior mean.  Performance was assessed by the sum of squared errors in estimating the true value of $\bbeta$.  Convergence of both algorithms was assessed by starting from multiple distinct points in $\mathbb{R}^p$ and checking that the final solutions were identical up to machine and/or Monte Carlo precision.  As before, for the classical bridge estimator, the regularization parameter $\nu$ was chosen by generalized cross validation; while for the Bayesian bridge, $\sigma$ was assigned Jeffreys' prior and $\nu$ a Gamma(2,2) prior.

\begin{table}
\begin{center}
\caption{\label{tab:estimationSSE} Average sum of squared errors in estimating $\bbeta$ for three different batches of 250 simulated data sets.}
\vspace{1pc}
\begin{tabular}{r r rrr}
			&& LSE & Bridge & Bayes \\
			\hline
$\alpha = 0.5$	&& 2254 & 1611 & 99 \\
$\alpha = 0.7$	&& 1994 & 406 & 225 \\
$\alpha = 0.9$	&& 551 & 144 & 85
\end{tabular}
\end{center}
\end{table}

Table \ref{tab:estimationSSE} shows the results of these experiments.  For all three choices of $\alpha$, the posterior mean estimator outperforms both least squares and the classical bridge estimator.  Sometimes the difference is drastic---such as when $\alpha = 0.5$, where the Bayes estimator outperforms the classical estimator by more than a factor of 16.

\section{Discussion}

This paper has described a series of tools that allow practitioners to estimate the full joint distribution of regression coefficients under the Bayesian bridge model.  Our numerical experiments have shown: (1) that the classical bridge solution, the posterior mode under a joint Bayesian model, and the posterior mean can often lead to very different summaries about the relative importance of different predictors; and (2) that using the posterior mean offers substantial improvements over the mode when estimating $\bbeta$ or making predictions under squared-error loss.  Both results parallel the findings of \citet{park:casella:2008} and \citet{hans:2008} for the Bayesian lasso.

The existence of a second, novel mixture representation for the Bayesian bridge is of particular interest, and suggests many generalizations, some of which we have mentioned.  Our main theorem leads to a novel Gibbs-sampling scheme for the bridge that---by virtue of working directly with a two-component mixing measure for each latent scale $\omega_j$---is capable of easily jumping between modes in the joint posterior distribution.  The evidence suggests that it is the best algorithm in the orthogonal case, but that it suffers from poor mixing when the design matrix is strongly collinear.  The chief limiting factor here is our inability to efficiently generate samples from the truncated multivariate normal distribution.  Luckily, in this case, the normal-mixture method based on the work of \citet{devroye:2009} for sampling exponentially tilted stable random variables performs well.    Moreover, both MCMC methods appear to alleviate some of the difficulties associated with slow mixing in global-local scale-mixture models described by \citet{hans:2008}, and further studied in the online supplemental file.   They also allow uncertainty about the concavity parameter to be incorporated naturally.  Both are implemented in the R package \verb|BayesBridge|, available through CRAN. Together, they give practioners a set of tools for efficiently exploring the bridge model across a wide range of commonly encountered situations.

\appendix

\section{Proofs}

\subsection{Theorem \ref{thm:kmonotone}}

\begin{proof}

Let $M_n$ denote the class of $n$-times monotone functions on $(0, \infty)$.  Clearly for $n \geq 2$, $f \in M_{n} \Rightarrow f \in M_{n-1}$.  Thus it is sufficient to prove the proposition for $k=n$. As the density $f(x)$ is symmetric, we consider only positive values of $x$.

The Schoenberg--Williamson theorem \citep[Theorems 1 and 3]{williamson:1956} states that a necessary and sufficient condition for a function $f(x)$ defined on $(0, \infty)$ to be in $M_n$ is that
$$
f(x) = \int_0^{\infty} (1-ut)_{+}^{n-1} \ dH(u) \, ,
$$
for some $H(u)$ that is non-decreasing and bounded below.  Moreover, if $H(u) = 0$, the representation is unique, in the sense of being determined at the points of continuity of $H(u)$, and is given by
$$
H(u) = \sum_{j=0}^{n-1} \frac{(-1)^j f^{(j)}(1/u) }{j!} \left( \frac{1}{u} \right)^j \, .
$$

Let $s = 1/u$.  This yields
\begin{eqnarray*}
f(x) &=& \int_0^{\infty} (1-x/s)_{+}^{n-1} \ dG(s) \\
G(s) &=& \sum_{j=0}^{n-1} \frac{(-1)^j f^{(j)}(s) }{j!} s^j \, .
\end{eqnarray*}
Now rewrite the kernel as a scaled beta density to give
$$
f(x) = \int_0^{\infty} \frac{1}{s} n \left(1-\frac{x}{s} \right)_{+}^{n-1} \ (s/n) dG(s) \, .
$$
Differentiating the CDF with respect to $s$ and absorbing the factor of $s/n$ into $G(s)$, we conclude that the mixing density is
$$
g(s) ds \propto n^{-1}  \sum_{j=0}^{n-1} \frac{(-1)^j }{j!} \left\{ j s^{j} f^{(j)}(s)   +  s^{j+1} f^{(j+1)}(s) \right\} \ ds\, ,
$$
and the result is proven.


\end{proof}

\end{spacing} 

\singlespace
\begin{footnotesize}
\bibliographystyle{abbrvnat}
\bibliography{masterbib,bayesbridge}
\end{footnotesize}

\end{document}